\documentclass[sn-mathphys,Numbered,11pt]{sn-jnl}% Math and Physical Sciences Reference Style
\usepackage{graphicx}%
\usepackage{multirow}%
\usepackage{amsmath,amssymb,amsfonts}%
\usepackage{amsthm}%
\usepackage{upgreek}%
\usepackage{subfigure}
\usepackage{mathrsfs}%
\usepackage[title]{appendix}%
\usepackage{xcolor}%
\usepackage{textcomp}%
\usepackage{manyfoot}%
\usepackage{booktabs}%
\usepackage{algorithm}%
\usepackage{algorithmicx}%
\usepackage{algpseudocode}%
\usepackage{listings}%
\theoremstyle{thmstyleone}%
\theoremstyle{thmstyletwo}%

\theoremstyle{thmstylethree}%

\begin{document}

\title[System performance of a TDM test-bed with long flex harness towards the new X-IFU FPA-DM]{System performance of a TDM test-bed with long flex harness towards the new X-IFU FPA-DM}

\author*[1]{D.~Vaccaro}\email{d.vaccaro@sron.nl}
\author[1]{M.~de~Wit}
\author[2]{J.~van~der~Kuur}
\author[1]{L.~Gottardi}
\author[1]{K.~Ravensberg}
\author[1]{E.~Taralli}
\author[3,4]{J.~Adams}
\author[4]{S.~R.~Bandler}
\author[4]{J.~A.~Chervenak}
\author[5]{W.~B.~Doriese}
\author[5]{M.~Durkin}
\author[5]{C.~Reintsema}
\author[3,4]{K.~Sakai}
\author[4]{S.~J.~Smith}
\author[3,4]{N.~A.~Wakeham}
\author[2]{B.~Jackson}
\author[1]{P.~Khosropanah}
\author[1,6]{J-R.~Gao}
\author[1,7]{J.W.A.~den~Herder}
\author[2]{P. ~Roelfsema}

\affil[1]{NWO-I/SRON Netherlands Institute for Space Research, 2333CA Leiden, The Netherlands}
\affil[2]{NWO-I/SRON Netherlands Institute for Space Research, 9747AD Groningen, The Netherlands}
\affil[3]{Center for Space Sciences and Technology, University of Maryland Baltimore County, Baltimore, MD21250 USA}
\affil[4]{National Aeronautics and Space Administration, Goddard Space Flight Center, Greenbelt, MD 20771, USA}
\affil[5]{National Institute of Standards and Technology, 325 Broadway, Boulder, CO, 80305-3328, USA}
\affil[6]{Optics Group, Delft University of Technology, Delft, 2628CJ, The Netherlands}
\affil[7]{Universiteit van Amsterdam, Science Park 904, 1090GE Amsterdam, The Netherlands}

\abstract{SRON (Netherlands Institute for Space Research) is developing the Focal Plane Assembly (FPA) for Athena X-IFU,  whose Demonstration Model (DM) will use for the first time a time domain multiplexing (TDM)-based readout system for the on-board transition-edge sensors (TES).  We report on the characterization activities on a TDM setup provided by NASA Goddard Space Flight Center (GSFC) and National Institute for Standards and Technology (NIST) and tested in SRON cryogenic test facilities.  The goal of these activities is to study the impact of the longer harness,  closer to X-IFU specs,  in a different EMI environment and switching from a single-ended to a differential readout scheme.  In this contribution we describe the advancement in the debugging of the system in the SRON cryostat,  which led to the demonstration of the nominal spectral performance of 2.8 eV at 5.9~keV with 16-row multiplexing, as well as an outlook for the future endeavours for the TDM readout integration on X-IFU's FPA-DM at SRON. }


\maketitle

\begin{quotation}
This paper is under publication in \textit{Journal of Low Temperature Physics}.
\end{quotation}

\section{Introduction}\label{intro}

The Advanced Telescope for High-Energy Astrophysics (ATHENA) is ESA's L2-class mission, slated to launch in 2037 into a L1 orbit between the Sun and the Earth to study the Hot and Energetic Universe.  It will feature two instruments: the Wide Field Imager (WFI) designed for surveys and the X-ray Integral Field Unit (X-IFU) \cite{xifu} optimised for imaging spectroscopy. The core of X-IFU will contain 1536 transition-edge sensors (TES) micro-calorimeters \cite{gottardites},  organized in an hexagonal array covering a field of view of 4' with a pixel size of $317~\upmu$m corresponding to 4'' angular resolution. The sensors will be cooled down to a base temperature of 50~mK by means of a multi-stage Adiabatic Demagnetization Refrigerator (ADR) and are designed to operate in the soft X-ray energy band, between 200~eV and 12 keV,  with a predicted energy resolution better than 2.5~eV for energies up to 7~keV.

The readout of hundreds (or thousands) of cryogenic sensors in a space-borne mission requires multiplexing in order to comply with heat load, mass and wire count limitations. For X-IFU, the chosen readout technology is time-division multiplexing (TDM) \cite{tdm}, a mature technology routinely used in ground-based X-ray and $\upgamma$-ray spectrometers, such as the SSRL beamlines at SLAC (USA), and recently in the Micro-X sounding rocket \cite{microx}.

In TDM readout, the TES are continuously dc-biased and are each coupled to a first stage SQUID (SQ1), activated one at a time via a a flux-actuated switch (FAS). SQ1 signals are amplified by a SQUID series array (SSA) and digitized by room temperature electronics. Each SQ1 represent a ``row" and the FAS are activated by sequential square wave signals, or row-addressing signals (RAS), generated by the digital readout electronics (DRE).  The SQ1s are hosted on ``MUX" chips, together with the shunt resistance $R_{sh}$ to voltage bias the TES and Nyquist inductors $L_{Ny}$ to limit the pixel bandwidth. Each row is active for a certain ``row time",  all the $N$ rows are readout within one ``frame" and the $M$ columns are readout in parallel. For X-IFU, a 48-row $\times$ 32-column TDM readout is envisaged, with row time of 160~ns.

SRON is developing the next iteration of X-IFU's Focal Plane Assembly Demonstration Model (FPA-DM), which will implement TDM readout.  For this reason,  the National Institute of Standard and Technology (NIST) and NASA Goddard Space Flight Center (USA) provided SRON with a complete TDM system to use as a test-bed towards the integration of TDM into the FPA-DM. In this paper the characterization activities of this system are reported.

\section{TDM cryogenic test-bed}

The TDM system is mounted in a Leiden Cryogenics (model CF-CS81-400) dilution refrigerator with a cooling power of 400~$\upmu$W at 120~mK. The setup consists of the components listed as follows.

\begin{figure}[h]%
\centering
\includegraphics[trim={0 0 0 6cm},clip,width=0.9\textwidth]{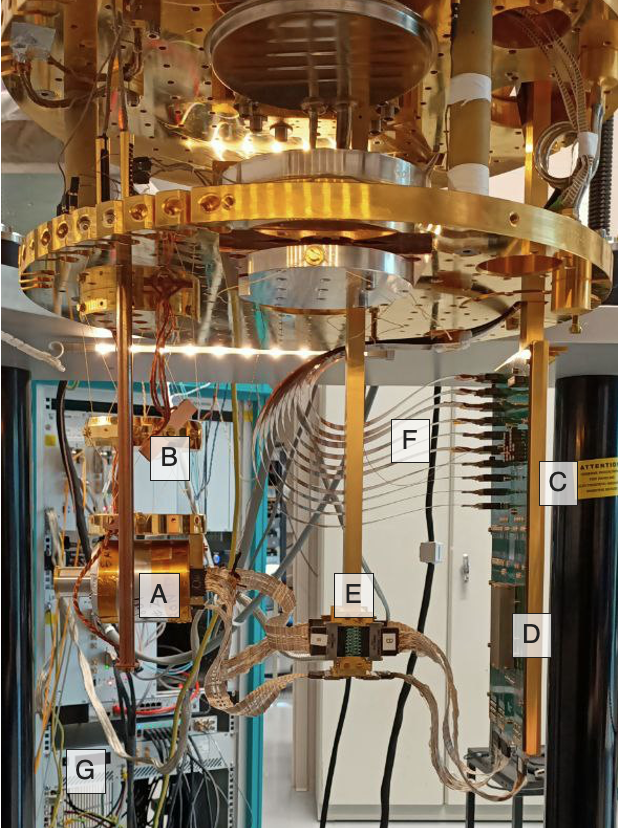}
\caption{Cryogenic components of the setup installed in the dilution refrigerator: 50 mK snout in Al superconducting shield (A), two-stage Kevlar suspension system (B), 4 K board (C), SSA SQUIDs (D), inter-thermal stage boards (E) for the NbTi braided looms connecting the 4 K to the 50~mK snout, flex harness (F) and in the background the DRE (G) mounted in an electronics rack.}\label{system}
\end{figure}

\begin{itemize}

\item A ``50~mK snout" hosting the TES and MUX chips ([A] in Figure \ref{system}). The detector chip is a kilo-pixel array with TES of 75$\times$75~$\upmu$m, $R_N = 9.55$~m$\Omega$, $T_C = 92.5$~mK and an average single-pixel energy resolution at a level of 2.3~eV at 6 keV, while the MUX chips are designed with $L_{Ny} = 640$~nH and $R_{sh} = 69\ \upmu\Omega$.  The snout is enclosed in an aluminum shell to shield the TES array and SQUIDs from the influence of external magnetic fields. An Iron-55 source is mounted on a chimney on the aluminum shield to illuminate the detectors with photons from the MnK$\upalpha$ line complex ($\simeq$~5.9~keV) to characterize the spectral performance.  A magnetic field can be applied locally via a superconducting coil placed below the TES array. The snout is mounted below the mixing chamber of the refrigerator using a two-stage Kevlar suspension ([B] in Figure \ref{system}) system to dampen mechanical vibrations \cite{gottardikevlar} and thermalized via copper braids. Temperature is monitored and controlled via a ruthenium oxide thermometer and a $750~\Omega$ heater installed on the snout's mechanical support.  A temperature of 50~mK is set, with a stability at a level of 1~$\upmu$K. 

\item A ``4~K board" ([C] in Figure \ref{system}) connected to the warm analog electronics (``Tower") via flex harness,  used to host the Series SQUID array (SSA, [D] in Figure \ref{system}).  The SSA chip is mounted inside a niobium casing for magnetic shielding, and is constituted by SA20a SQUIDs manufactured at NIST. The board is mounted on a mechanical support located below the mixing chamber, but thermalized to the refrigerator's 4~K plate via a copper rod passing through the clearshots of the refrigerator's plates. This extension from the far 4~K plate is necessary to limit the distance between the SSA and the snout, so that there is sufficient interstage bandwidth between the two SQUID stages. Because of the weight of the board's mechanical support and the length of the rod, a Kevlar support structure on top of the mixing chamber has been devised to dampen mechanical vibrations. Communication between the 4~K board and the snout is done via NbTi wiring, with a thermal break achieved through inter-stage boards thermalized at the heat exchanger plate ($\sim$ 100~mK) ([E] in Figure \ref{system}).

\item Flex harness ([F]  in Figure \ref{system}) connecting the 4~K board to the ``Tower". There are a total of 10 flexes,  each constituted by 8 Cu single-ended signal lines.  \iffalse Each line is composed of a signal trace and a return trace, both $\approx 0.152$~mm wide and with a pitch of 2.54~mm. Each neighbouring trace pair is separated by a 0.127~mm polymide film ($\epsilon_r$~=~4 at 1 MHz).  \fi The total length of the harness is 1.54~m. The flexes are stacked and thermalized via dedicated heatsink plates at 4~K and 50~K. 

These flexes differ from the ``legacy" design  used in standard TDM systems,  either in length ($\sim$43~cm), trace geometry (wider return traces) and material (Cu instead of CuNi).  The motivation was the reproduction of the nominal performance of a TDM system in our dilution refrigerator, where the distance between 4~K and room-temperature is larger than in the ADRs available at NIST and NASA-GSFC. Furthermore,  it was also an important and necessary step to characterize the system with longer harness, closer to the current 2.3~m required for X-IFU. In order to maintain the typical ``row time" of 160~ns, these ``SRON" flexes were dimensioned at NIST to keep the electrical resistance below 20~$\Omega$ to be compliant with the requirements for the electronics, as well as limiting the capacitance between signal and return lines, keeping it at the same level of the ``legacy" flexes, to prevent bandwidth loss\iffalse (current requirement is approximately 12 MHz)\fi. This design however resulted in a larger cable inductance, which enhanced electrical crosstalk in the system,  ultimately affecting the system performance already in early tests performed at NIST, as discussed in more detail in the next Section. 

\item The ``Tower" (NIST) is the warm front end electronics (WFEE) connected via the flexes to the 4~K PCB,  by a vacuum feed-through on top of the cryostat.  The Tower is composed of modular boards, each assigned to a specific function: amplification of the signal coming from the SSA via a low-noise amplifier (LNA), biasing via DAC circuitry to the quiescence operating point of detectors, SQ1s and SSA and connection to the feedback circuitry of the digital readout electronics. Control of the bias boards is performed through a serial protocol, instantiated from the measurement computer via a optical fiber connection. Connection to the digital readout electronics is performed by HDMI cables for the RAS and BNC cables for ADC and feedback.

\item The DRE, Digital Readout Electronics (NASA-GSFC), is composed of two elements: the Column Box and the Row Box ([G] in Figure \ref{system}).  The DRE has the important function of maintaining the synchronicity among the various signals, done via the generation of several clocks (frame clock, line clock) derived from a master clock of 245.76 MHz. Besides timing control, the DRE generates the RAS to switch the multiplexers (Row Box), the digital feedback for the flux-locked loop (FLL) of the SQ1s and DAQ back-end for data streaming (Column Box). More details on this DRE can be found in \cite{kazudre}.
\end{itemize}

The system is designed to host up to 8 readout channels. In the case of the system received by NASA GSFC, only 2 channels are populated and the available DRE allows a single channel (or Column) readout. Therefore, in this paper the focus is on the characterization of one single channel. The Row Box allows for a readout of up to 40 rows,  however the MUX chips in the snout are designed for up to 32-row multiplexing. Of these 32, 2 rows are disconnected at SQ1 level due to defects in the chip. Furthermore, 3 rows were observed to be extremely noisy, likely due to defects in the flex harness, so that in the considered channel there are 27 rows which exhibit the expected SQUID and TES response.

\section{Characterization and performance}

\subsection{EMI optimization}

To prevent ground loops,  a ``star" grounding scheme is used, where the top of the cryostat is used as ground reference. All measurement electronics are grounded via thick copper braids to such reference.  Electronics modules, including the Column and Row Boxes, are powered on via an isolation transformer to decouple the grounding of the TDM to the ground of the wall socket. The top of the cryostat is then connected to a ``clean" ground reference in the laboratory via a thick copper grounding cable. To electrically decouple the connection between the Tower and the Column box to the measurement computer, optical fibers are used.

\begin{figure}[!h]%
\centering
\includegraphics[width=0.7\textwidth]{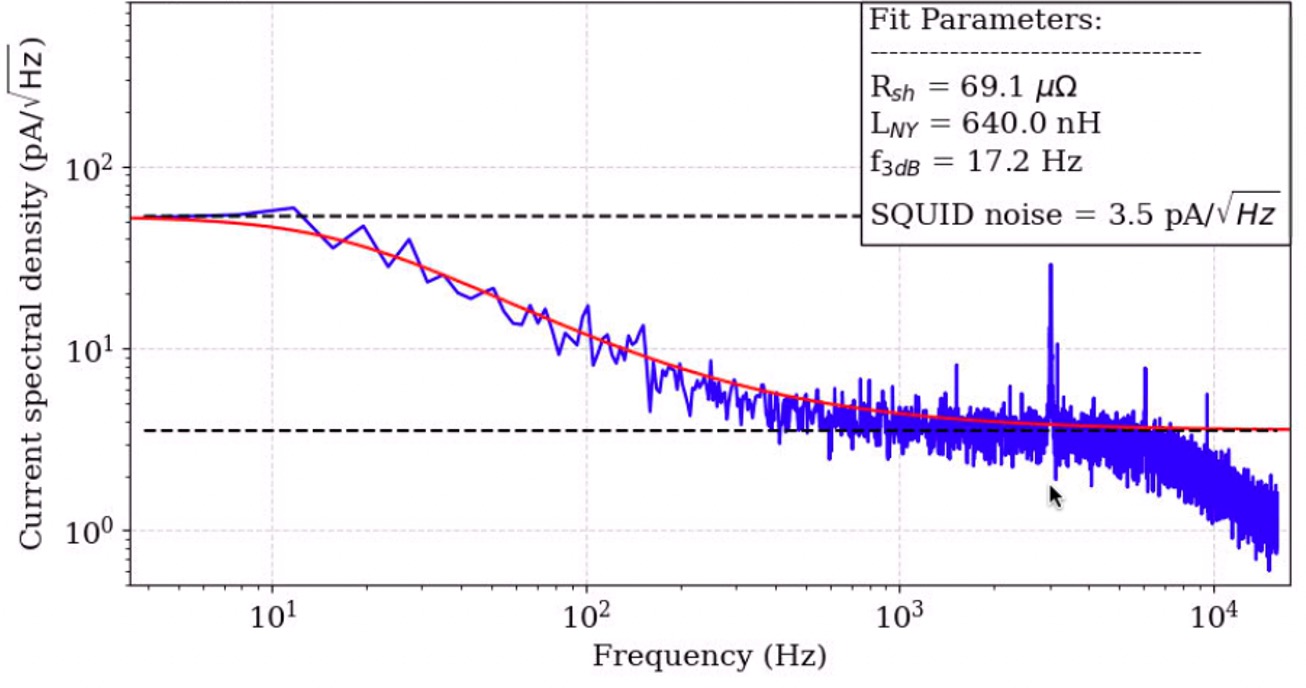}
\caption{Measured SQUID noise for a single pixel (blue curve) and fit (red curve) with $R_{sh}$ and $L_{Ny}$ fixed to the nominal value, after EMI optimization. The noise level is measured with the TES in superconducting state, by averaging the FLL signal between 1.9 kHz and 2.1 kHz, outside the Johnson noise cutoff $R_{sh}/L_{NY}$ of the shunt resistance. }\label{squidnoise}
\end{figure}

Despite this optimization, strong noise lines at 50 Hz and higher harmonics have been initially observed in SQUID noise measurements. This was identified as due to electromagnetic interference (EMI) in the loop constituted by the $\sim$~3~m HDMI and BNC cables connecting the DRE to the Tower. In fact, it was observed that bundling and intertwining the cables together, to minimize the area of the loop and reject common-mode noise, significantly reduced the amplitude of such noise lines.  Finally, enclosing the cable bundle in aluminium foil completely removed the 50 Hz noise lines.  High-frequency (several kHz) lines, thought to be spurs from the electronics, remained present in the system, as depicted in Figure \ref{squidnoise}.  However, their frequency is sufficiently higher with respect to the thermal bandwidth of the detectors to consider their impact negligible. Note that the measurement was performed with the TES in superconducting state, so that the Johnson noise from the shunt resistor and the $R_{sh}/L_{NY}$ roll-off could be observed. The measured values correspond very well to the expected values, verifying the cold readout circuit as well as thermometry

\subsection{Impact of magnetic field}\label{bmeas}

For optimal spectral performance, the behaviour of the TES under external magnetic field has to be studied, since the best detector operation is achieved when no perpendicular magnetic fields are present. These fields might be caused by external sources such as pulse tubes, Earth magnetic field, EMI from laboratory equipment, etc.  Furthermore, TES under dc-bias suffer from the effects of self-field, $i.e.$ the magnetic field generated by the loop made by the leads and the TES itself \cite{smithb}. Such magnetic field adds to the residual field at the TES array and is bias point dependent. Such magnetic fields are typically compensated with a superconducting coil mounted just below the TES array,  a configuration also envisaged for X-IFU.

\begin{figure}[h]%
\centering
\subfigure[]{\includegraphics[width=0.49\textwidth]{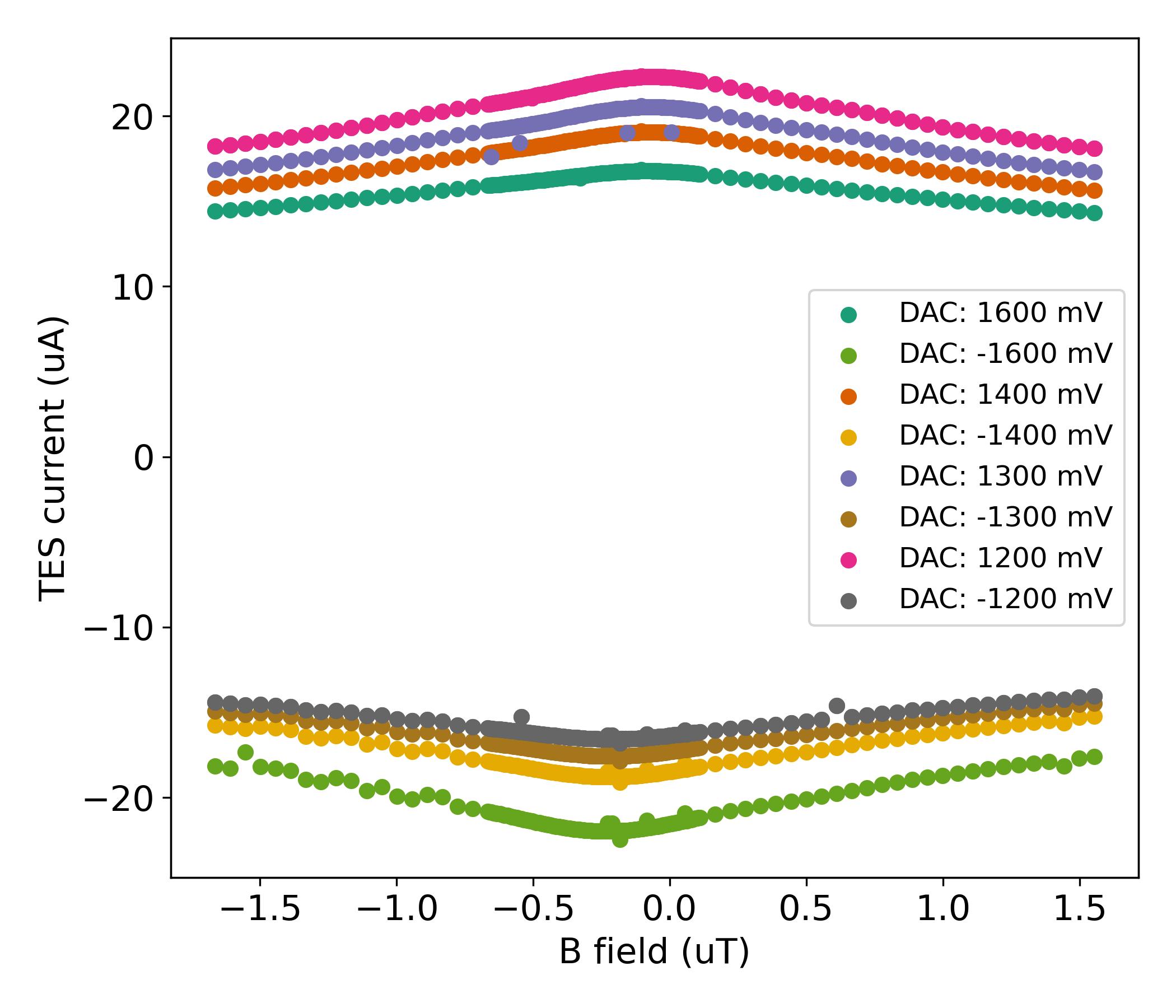}}
\subfigure[]{\includegraphics[width=0.49\textwidth]{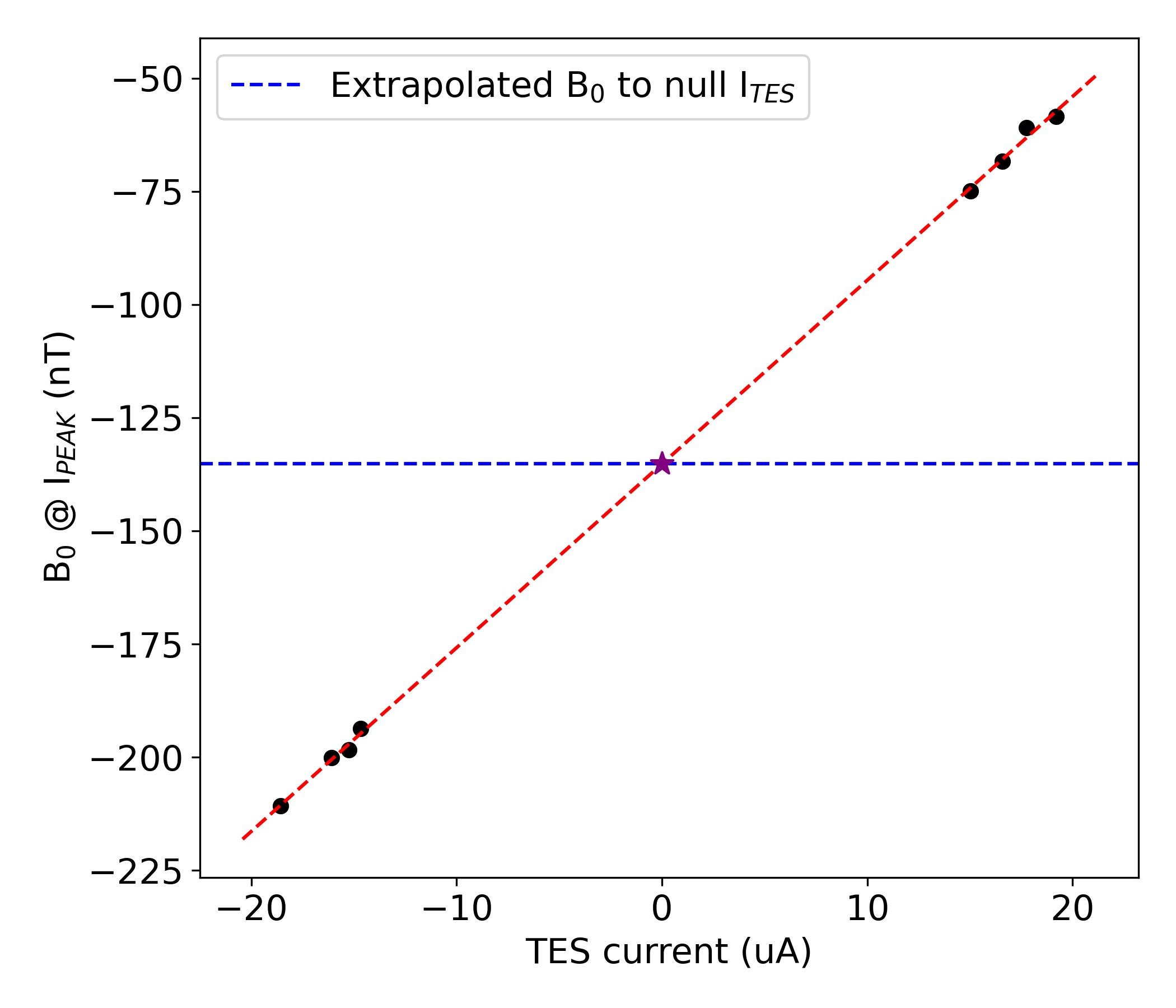}}
\caption{Example of $I_{TES}$ vs $B_{\text{ext}}$ as a function of bias point (a) and fit of $B_0$ vs $I_{TES}$ to estimate the cancelling field at zero TES current, where no self-field effect is present.}\label{ivsb}
\end{figure}

It was assumed that for best spectral performance an external magnetic field equal to the residual field without self-field is used. To estimate this, TES IV curves at different magnetic fields $B_{\text{ext}}$ are measured to extract how the calibrated TES current $I_{\text{TES}}$ changes as a function of applied field for a given bias point. We then obtain $I_{\text{TES}}$ versus $B_{\text{ext}}$ curves, showing the well-known Fraunhofer-like trend. The peak of such curve for different bias points can be used to extrapolate the cancelling field corresponding to when no current is flowing through the TES, as shown in Figure \ref{ivsb}.

As a sanity check, we also estimated the optimal cancelling field with an alternative method, raising the bath temperature just below $T_C$ and measuring the $I(B)$ curve with the minimum bias voltage to bring the TES in transition. In this configuration,  $I_{\text{TES}}$ is low and self-field can be considered negligible, so that the peak of the measured Fraunhofer pattern can be assumed to be purely due to the residual magnetic field.

We found the average of the cancelling fields measured with both methods to be consistent.  We measured an average cancelling magnetic field at a level of 150~nT with no evident gradient across the array,  comparable with what typically measured for other TES setups readout under Frequency Domain Multiplexing hosted in the same cryostat \cite{fdm}. The absence of a magnetic field gradient is important for flight operation, particularly for TDM readout since (1) dc-biased pixels are more sensitive to magnetic fields than ac-biased pixels \cite{fdmbsens,lgss} and (2) each TES shares the same bias voltage and can't be tuned individually to compensate for eventual different magnetic field offsets across the array.

\iffalse

\begin{figure}[h]%
\centering
\includegraphics[width=0.6\textwidth]{Bfield_map.jpg}
\caption{Map of the canceling magnetic field measured for TES in one Column.}\label{bfieldmap}
\end{figure}

\fi

\subsection{Row time tuning}

As mentioned in Section~2, the harness connecting the 4 K board to the Tower was redesigned for this system to maintain the electrical resistance and capacitance at the same level as in the ``legacy" flex, given the $\approx 4\times$ length of the flexes with respect to standard NIST systems. This design however produced an unforseen predominantly inductive crosstalk.

\iffalse

\begin{figure}[h]%
\centering
\includegraphics[width=0.8\textwidth]{Ringing.jpg}
\caption{Ringing (a) as a function of input low-pass filter (LPF) order at the ADC (b), sampled over one frame, $i.e.$ 58.6 $\upmu$s, with system parameters as in the 3rd column of Table 1.  Increasing the order of the filter concurrently smooths the edges of the RAS square wave, effectively reducing the sampling window to construct the error signal. With 366~ns row time, a 5th order results in a too large smoothing of the square wave for practical use: for our measurements we therefore used a 4th order filter (purple curve),  as the best trade-off available between dampening of the ringing and sharpness of the row-addressing signal square-wave.}\label{ringing}
\end{figure}

\fi

\begin{figure}[h]%
\centering
\subfigure{\includegraphics[width=0.49\textwidth]{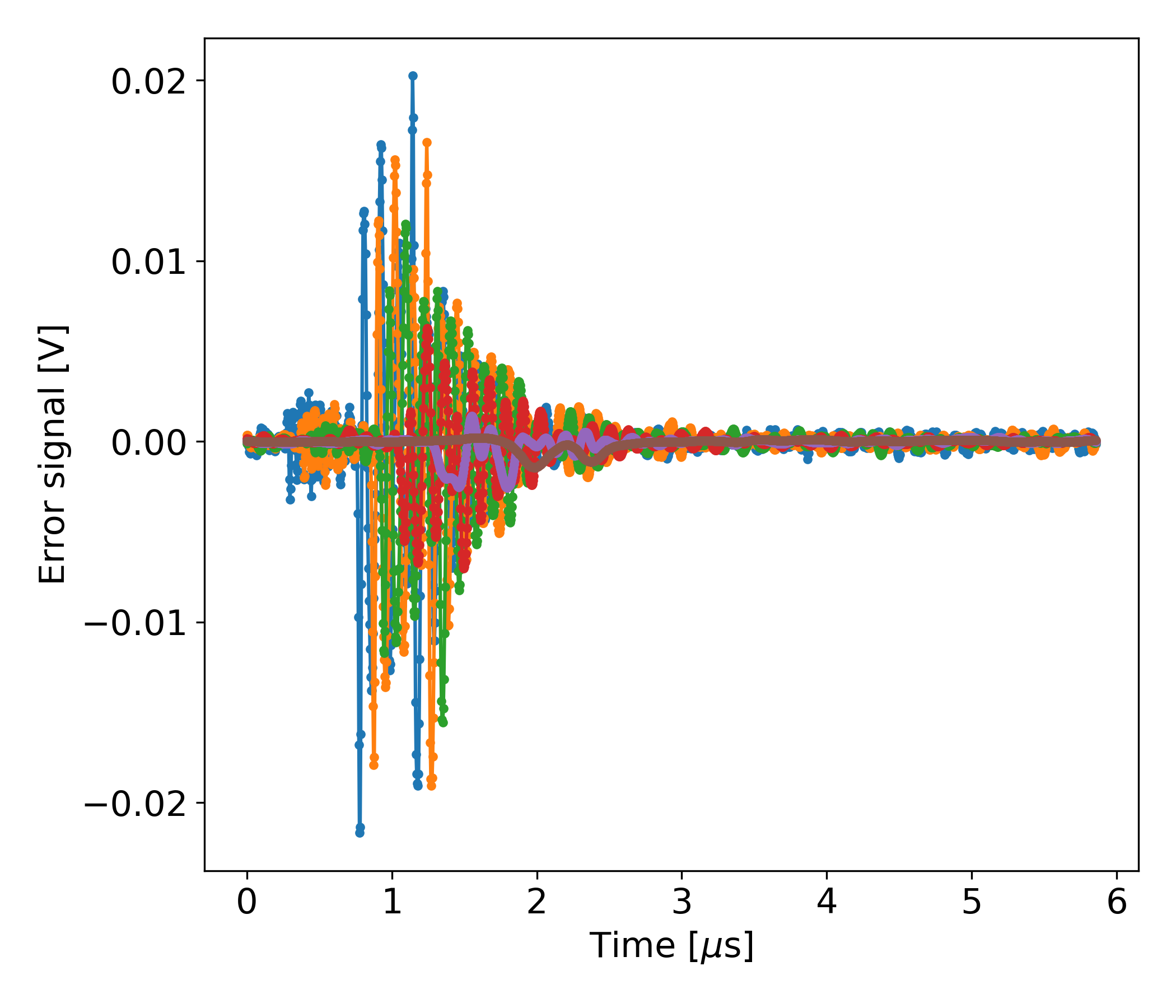}}
\subfigure{\includegraphics[width=0.49\textwidth]{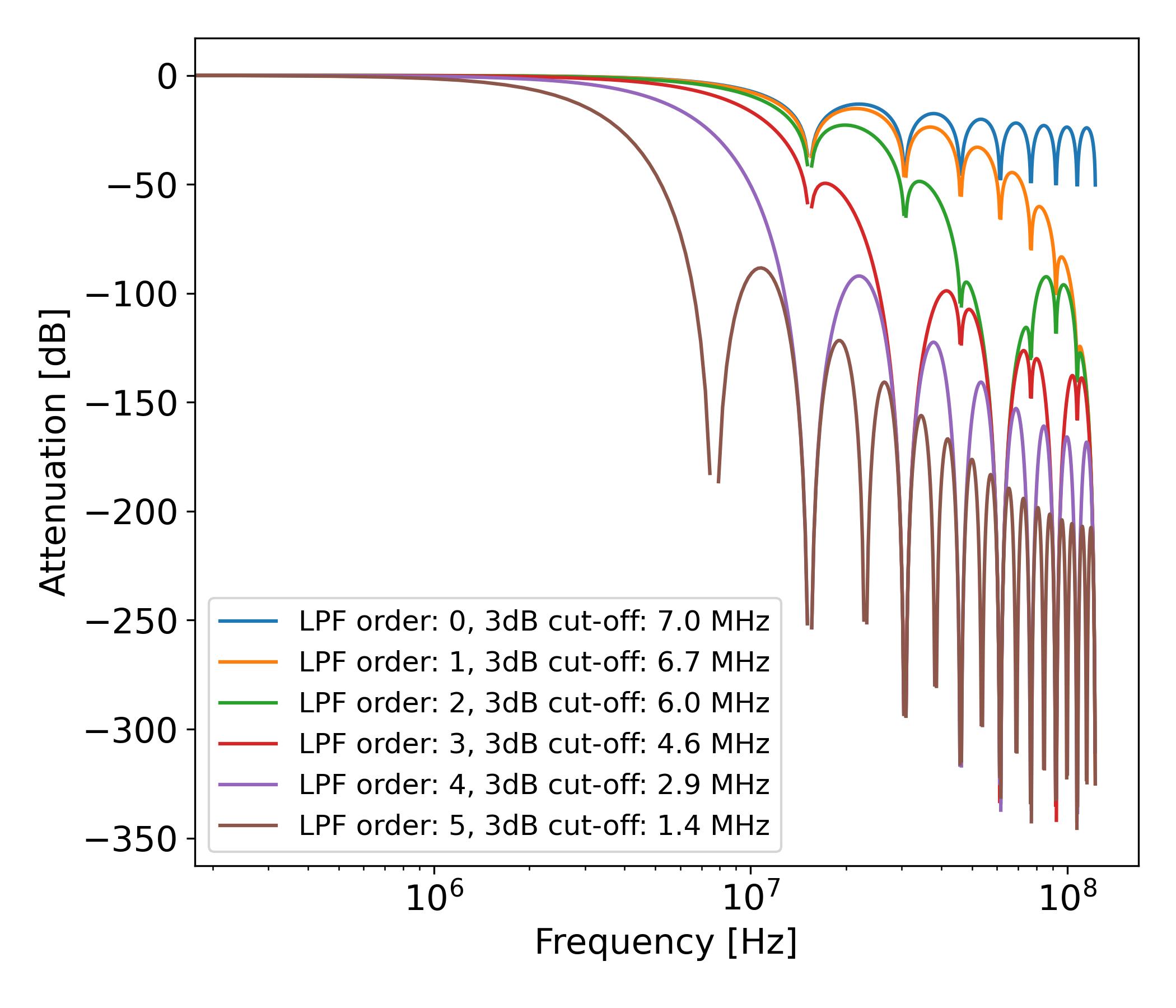}}
\caption{Ringing (left panel) as a function of input low-pass filter (LPF) order at the ADC, sampled over one frame, $i.e.$ 58.6 $\upmu$s, with system parameters as in the 3rd column of Table 1.  The calculated transfer function of the digital LPF for 16 samples is shown in the right panel. Increasing the order of the filter concurrently smooths the edges of the RAS square wave, effectively reducing the sampling window to construct the error signal. With 366~ns row time, a 5th order filter limits the bandwidth too much, resulting in an excessive smoothing of the square wave for practical use: for our measurements we therefore used a 4th order filter (purple curve),  as the best trade-off available between dampening of the ringing and sharpness of the row-addressing signal square-wave.}\label{ringing}
\end{figure}

This crosstalk manifests as underdamped oscillations (or ringing) in the electrical signals in the system. They are expected to be initiated by two mechanisms: the row-addressing signals activating the TDM rows and change in the feedback signals, due $e.g.$ to an X-ray pulse. The latter ringing is observable, for example by injecting a current-jump in the system equivalent to 1 $\Phi_0$ at SQ1 feedback coil, to mimic a X-ray pulse.  In this fashion, we measured a peak frequency for this oscillations at a level of 8-9 MHz (Figure \ref{ringing}), well within the required $\sim 12$~MHz system readout bandwidth.  In standard TDM systems with shorter harness, this oscillation is also present but decays fast enough to be outside of the readout bandwidth, for the same timing parameters.

The nominal spectral performance for this system was at a level of 2.8~eV at energies of 6~keV with 32-row readout and ``legacy" flexes.  Because of the larger crosstalk with the longer flexes, a row time of 160~ns did not allow for a sufficient decay of the oscillation before signal sampling, resulting in severe performance degradation.  Increasing the row time while keeping untouched the multiplexing factor would have resulted in an increased frame time and hence undersampling of the X-ray pulses. To maintain the same frame rate while increasing the row time,  the multiplexing factor must then be reduced.  In tests with the longer ``SRON" flexes, NIST demonstrated that multiplexing 16 rows instead of 32 allowed to set a row time of 360 ns,  with a settling time of 224 ns and a sampling time of 136 ns. These settings were chosen to wait for a sufficient damping of the ringing and then to integrate the ADC signal over one period,  to effectively average out the oscillation.  The choice of these parameters allowed to recover the expected performance in multiplexing of 2.8~eV with a reduced number of rows,  from 32 to 16, as a consequence of extending the row time from the nominal 160~ns. This configuration was considered as the reference to reproduce with the system installed in the SRON cryostat.

\subsection{X-ray spectral performance}

First X-ray measurements with the TDM system in the SRON cryostat were performed in a 4-row multiplexing configuration,  leaving sufficient margin in the timing parameters to consider these measurements as an acceptable proxy for single pixel measurements. 

To estimate the spectral performance, about 5000 X-rays per pixel were collected from an Iron-55 source, emitting fluorescence photons corresponding to the Mn-K$\upalpha$ line complex (5.9 keV). The energy of the X-ray events is evaluated using the optimal filtering technique. The spectral performance is then calculated by fitting the Mn-K$\upalpha$ model \cite{holzer} to the collected events by minimizing the Cash statistics in the maximum-likelihood method \cite{cstat}.

The measurements consisted of collecting X-ray photons at different TES bias points, in the range of $R/R_N$ between 0.25 and 0.10. The best energy resolutions were at a level of 2.4~eV, measured for $R/R_N \approx 0.18$, consistent with what measured at NIST in single-pixel mode before shipping of the system to SRON.  The energy resolution as a function of external magnetic field was also measured, finding that the best values corresponded to a cancelling field of 150~nT, consistent with expectations from the tests reported in Subsection \ref{bmeas}.

\begin{table}[h]
\begin{tabular}{@{}lllll@{}}
\toprule
& \textbf{NIST 2019 demo} \cite{tdm2019} & \textbf{NIST} & \textbf{SRON}  & \textbf{X-IFU nominal}\\
\midrule
\textbf{Master Clock}    & 125 MHz & 125 MHz   & 245.76 MHz & 125 MHz\\
\textbf{N-row}       & 40 & 16 & 16 & 48\\
\textbf{Line rate}  & 6.25 MHz & 2.78 MHz & 2.73 MHz & 6.25 MHz\\
\textbf{Frame rate} & 156.25 kHz & 173.61 kHz & 170.76 kHz & 130.21 kHz\\
\textbf{Row time}  & 160 ns & 360 ns   & 366 ns & 160 ns\\
\textbf{Flex lenght} & 43 cm & 154 cm & 154 cm & 230 cm\\
\textbf{Sampling time} & 32 ns & 136 ns & 65 ns & 32 ns\\
\textbf{Settling time}   & 128 ns & 224 ns   & 301 ns & 128 ns\\
\textbf{dE$_{\text{sum}}$} & 2.2 eV & 2.8 eV & 2.8 eV & 2.5 eV \\
\botrule
\end{tabular}
\caption{Comparison of timing parameters for different TDM demonstrations: first column is a reference for typical operation in NIST systems; second column are the parameters set at NIST for the tests of the system object of this paper (longer flex harness included); third column are the parameters (with the DRE from NASA GSFC) used to replicate the results obtained at NIST; fourth column are the required or expected parameters for Athena X-IFU. }\label{tab1}
\end{table}

\begin{figure}[h]%
\centering
\includegraphics[width=0.9\textwidth]{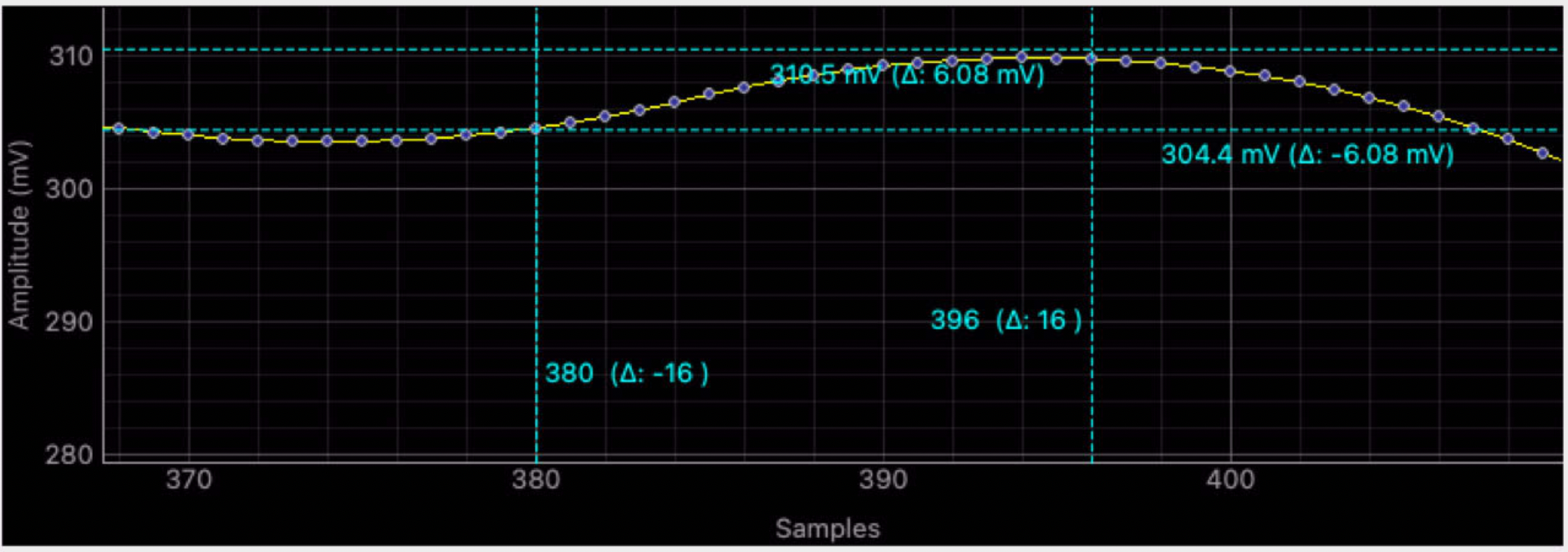}
\caption{Visualization on the digital scope for the fine tuning of the position of the 65 ns (16 samples) integration window on the RAS for the generation of the error signal.  The curve depicted here is a zoom on the crest of the square wave,  visibly oscillating due to the presence of the ringing.  This is the configuration with an input LPF of 4th order.  Using a 5th order LPF would result in this ``crest" being reduced to only few samples.}\label{timing}
\end{figure}

For 16-row multiplexing measurements however, due to design differences of the DRE from NASA GSFC received for the system at SRON with respect to the NIST electronics,  the exact same timing parameters could not be set. In particular, we could not match the 136~ns sampling time, since the maximum number of integration samples in the DRE is 16, so that $16/245.76\ \text{MHz} \approx 65$~ns. For this reason, the approach to average out the remnant of the ringing over one period of oscillation is not feasible. 

Setting the same row time and settling time with a limit of 65~ns sampling time yielded a summed spectral performance at a level of 3.1~eV at 5.9 keV.  To circumvent this limitation, we made use of the additional $\approx$~70~ns available,  increasing the settling time and moving as far away as possible the 65~ns integration window, up to right before the falling edge of the square wave smoothed by the 4th order low-pass filter (see Figure \ref{timing} for a visual representation). This approach successfully minimized the impact of the oscillation and allowed to retrieve the nominal system performance.  For reference, a comparison of timing parameters for a standard TDM system, the tests for this system at NIST and SRON, and the nominal parameters for X-IFU is reported in Table \ref{tab1}. 

\begin{figure}[h]%
\centering
\includegraphics[width=0.9\textwidth]{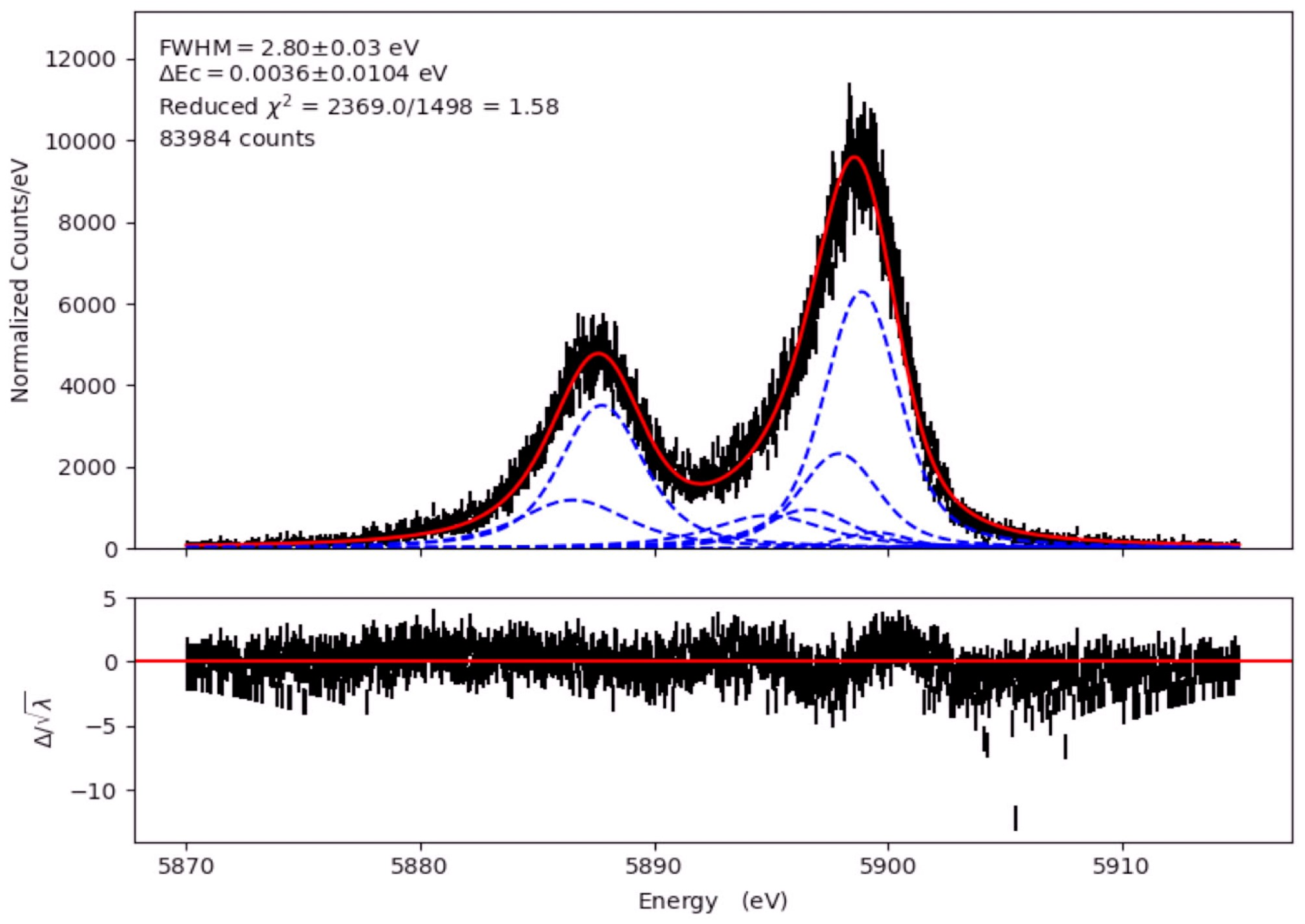}
\caption{Summed spectral performance measured with 16-row TDM multiplexed readout, for TES setpoint $R/R_N \approx 0.18$.  The histogram for the experimental data is presented in black. The best fit of the data with a Voigt function is presented in red. The single lines whose convolution represents the model for spectral complex are presented in blue.}\label{16mux}
\end{figure}

Using this configuration, we repeated the tests at different TES set-points and we measured for two different bias voltages ($R/R_N$ between 0.17 and 0.18) an energy resolution of $2.80 \pm 0.03$~eV at 5.9~keV. Figure~\ref{16mux} shows one of such spectra, consistent to what measured at NIST before delivery of the system to SRON.

The system was tested also with higher multiplexing factors.  With 32-rows, setting the row time to the X-IFU nominal 160~ns, the system could not maintain a stable FLL. An operative 24-row configuration was achieved, with a stable FLL: as expected, in this case there was no margin to sufficiently mitigate the impact of the crosstalk, and summed spectral performance was at a level of 7~eV.

\section{Summary and future outlook}

We reported on the characterization and test in SRON cryogenic facilities of a cold test-bed for a TDM readout provided by NIST and NASA GSFC towards the new FPA-DM for Athena X-IFU. The integration of the system in the SRON dilution refrigerator required harness between the warm front-end electronics and the cold SQUID amplifier with a length four times larger than what used in standard NIST systems (1st column in Table 1). This resulted in a high level of electrical crosstalk, posing the necessity to reduce the multiplexing factor from 32 to 16 to sufficiently relax the row time, from 160~ns to 360~ns: this allowed to increase the settling time and integration time, to 224~ns and 136~ns respectively,  matching the oscillation period to average it out. 

In these tests, performed with NASA GSFC digital readout electronics, a further challenge consisted in the upper limit of 65~ns on the sampling time. Fine tuning the position of the sampling window allowed to minimize the remnant impact of the ringing: in this way we could measure a spectral performance of $2.80 \pm 0.03$~eV at 5.9~keV with 16-row multiplexing, thus reproducing the system performance measured by NIST.

A new flex with an improved design is currently being implemented,  to further mitigate the crosstalk effects and retrieve the nominal 32-row multiplexing factor with no degradation in energy resolution.  Furthermore, the switch from single-ended to differential readout is expected to significantly reduce the magnitude of the ringing on the system, combined with active LNA input termination: its implementation requires a redesign of mainly harness and warm electronics, which is currently being pursued both by NIST and the Centre National de la Recherche Scientifique (France), the latter being responsible for the delivery of the WFEE for X-IFU.  Preparations are ongoing to demonstrate experimentally that the 160~ns row time, to date only achieved with the shorter ``legacy" flexes,  will be maintained with the proposed expedients for the $\gtrsim2$~m harness required for X-IFU.

\section*{Acknowledgements}

SRON is financially supported by the Nederlandse Organisatie voor Wetenschappelijk Onderzoek.
This work is part of the research programme Athena with project number 184.034.002, which is (partially) financed by the Dutch Research Council (NWO).

\section*{Data availability}

The corresponding author makes available the data presented in this paper upon reasonable request.

\end{document}